\documentclass[10pt,letterpaper]{article}
\usepackage{opex3}
\usepackage{amsmath, mathtools}
\usepackage{amsfonts}

\begin{document}

\title{Photoelastic coupling in gallium arsenide optomechanical disk resonators}

\author{Christopher Baker, William Hease, Dac-Trung Nguyen, Alessio Andronico, Sara Ducci, Giuseppe Leo, and Ivan Favero$^*$}

\address{Universit\'{e} Paris Diderot, Sorbonne Paris Cit\'{e}, Laboratoire Mat\'{e}riaux et Ph\'{e}nom\`{e}nes Quantiques,
CNRS-UMR 7162, 10 rue Alice Domon et L\'{e}onie Duquet, 75013 Paris, France
}

\email{$^*$ivan.favero@univ-paris-diderot.fr} 



\begin{abstract}
We analyze the magnitude of the radiation pressure and electrostrictive stresses exerted by light confined inside GaAs semiconductor WGM     optomechanical disk resonators, through analytical and numerical means, and find the electrostrictive force to be of prime importance.
We investigate the \emph{geometric} and \emph{photoelastic} optomechanical coupling resulting respectively from  the deformation of the disk boundary and from the strain-induced refractive index changes in the material,  for various mechanical modes of the disks.
Photoelastic optomechanical coupling is shown to be a predominant coupling mechanism for certain disk dimensions and mechanical modes, leading to total coupling g$_{om}$ and g$_0$ reaching respectively 3 THz/nm and 4 MHz.
Finally, we point towards ways to maximize the photoelastic coupling in GaAs disk resonators, and we provide some upper bounds for its value in various geometries. 
\end{abstract}



\bibliographystyle{osajnl}
\bibliography{biblioOE} 


\section{Introduction}

The field of optomechanics \cite{marquardt2009optomechanics, favero2009optomechanics, aspelmeyer2013cavity}
offers a rich array of applications spanning mechanical ground-state optical cooling \cite{teufel2011sideband, chan2011laser}, force and acceleration sensing \cite{forstner2012cavity, krause2012high}, wavelength conversion \cite{notomi2006optomechanical, liu2013electromagnetically} and all-optical tuning of photonic circuits \cite{deotare2012all,fong2011tunable,van2010optomechanical}.
In this context, semiconductor optomechanical disk resonators \cite{ding2010high, sun2012high, Jiang12, xiong2012aluminum}
are of particular interest due to their high optical quality factors (Q) and ability to confine both optical and mechanical energy in a reduced ($\sim \lambda^3$) interaction volume, thus providing  very strong optomechanical coupling. Alongside Silicon (Si), 
GaAs is a  platform of great potential for integrated photonics, as it allows for the integration of high optical Q and GHz high mechanical Q resonators \cite{usami2012optical, nguyen2013ultrahigh} with strong optomechanical coupling  \cite{ding2011wavelength} directly on-chip \cite{baker2011critical}. The GaAs platform furthermore enables the addition of electrically driven optically active elements, as well as the inclusion of quantum dots or quantum wells \cite{peter2005exciton} offering novel hybrid  optomechanical coupling schemes \cite{PhysRevLett.112.013601}.

\begin{figure}
\centering
\includegraphics[width=.65\textwidth]{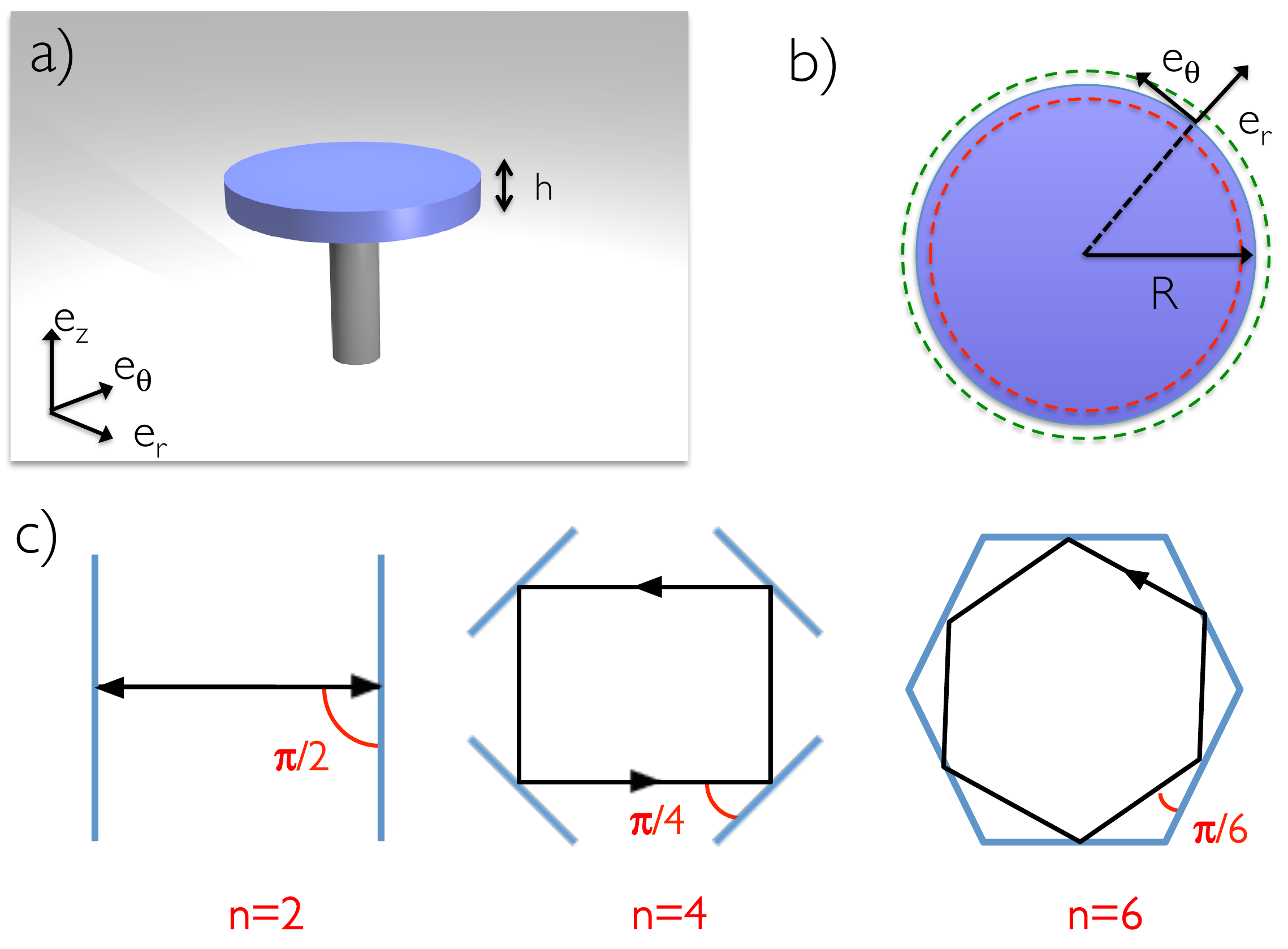}
\caption{(a) Schematic side-view of a GaAs disk of thickness $h$ (blue), positioned atop an AlGaAs pedestal (grey), along with the cylindrical coordinates used throughout this work. (b) Top view of a GaAs disk of radius R (blue). The dashed green and red lines represent the radial deformation of the disk by a mechanical mode. (c) Schematic view of an optical cavity composed of n=2, 4 or 6 mirrors, and the associated grazing angles.}
\label{figintro}
\end{figure}

The optomechanical resonators described in this work are composed of a micrometer-sized GaAs disk, isolated from the sample substrate atop an Aluminum Gallium Arsenide (AlGaAs) pedestal (Fig.\ref{figintro}(a)). The GaAs disk supports high Q optical WGMs located on the periphery of the disk, which are identified by their radial order $p$ and azimuthal number $m$ \cite{ding2010high, andronico2008difference}. The disk also supports a variety of in- and out-of-plane mechanical modes \cite{ding2010high}. A radial contour mechanical mode is schematically depicted in Fig. \ref{figintro}(b).

The photons confined inside the semiconductor disk exert two different stresses which will be detailed in the following: a \emph{radiation pressure} ``pushing the walls of the optical cavity apart'' and an \emph{electrostrictive} stress linked to the material's photoelasticity. Recently, Rakich et al. showed that for certain geometries of straight silicon photonic waveguides the electrostrictive stress could be commensurate with the radiation pressure stress commonly studied in optomechanics \cite{rakich2010tailoring, rakich2012giant}. In this paper we study the magnitude of these optical stresses in GaAs optomechanical disk resonators. We investigate the associated \emph{geometric} and \emph{photoelastic} optomechanical coupling strengths, resulting respectively from  the deformation of the disk boundary and from the strain-induced refractive index changes in the material, for various mechanical modes of the disk. We propose different computational methods, from analytical models leading to useful scaling formula, to full numerical approaches providing precise values of the coupling strength as a function of the mechanical mode and of the disk radius. For certain  mechanical modes, photoelasticity is a predominant optomechanical coupling mechanism, resulting in total coupling strengths
g$_{om}$ and g$_0$ that reach respectively 3 THz/nm and 4 MHz.
Finally we propose some simple rules to maximize the value of this coupling.

\subsection{Optomechanical coupling}
The disk resonator is described by the standard optomechanical Hamiltonian $\hat{H}$ describing an optical field coupled to a mechanical resonator \cite{aspelmeyer2013cavity}:
\begin{equation}
\hat{H}=\hbar \omega_0 \hat{a}^\dagger \hat{a}+ \hbar \Omega_M \hat{b}^\dagger \hat{b}\,- \hbar g_0 \hat{a}^\dagger \hat{a}\left(\hat{b}^\dagger + \hat{b}\right) 
\end{equation}
with $\omega_0$ and $\Omega_M$ respectively the optical and mechanical angular resonance frequency and $\hbar$ the reduced Planck constant.  $\hat{a}^\dagger$ ($\hat{b}^\dagger$) and $\hat{a}$ ($\hat{b}$) are respectively the photon (phonon) creation and annihilation operators. The optomechanical interaction can be defined in terms of the optomechanical coupling strength $g_{om}=-\frac{\mathrm{d} \omega_0}{\mathrm{d} x}$, representing the shift in the optical resonance frequency for a given mechanical displacement $dx$ or, in a complementary way, by $g_0=g_{om} x_{\mathrm{ZPF}}$, which represents the optical frequency shift for a mechanical displacement equal to the zero point fluctuations $x_{\mathrm{ZPF}}$. For completeness, we will quote both g$_{om}$ and g$_0$ in this work, fixing $x$ to be the maximum amplitude of displacement of the resonator \cite{ding2010high}.

\subsection{Mechanisms of optomechanical coupling}
A confined optical wave in the disk is only resonant provided it closes upon itself in phase after a round-trip, respecting the condition:
$2\pi n_{\mathrm{eff}} R \simeq m \lambda$, with $n_{\mathrm{eff}}$ the WGM effective index, $\lambda$ the optical freespace wavelength, R the disk radius and $m \in \mathbb{N}$. From this it appears that the resonance wavelength is modified by a small mechanical displacement $dx$ that changes the cavity radius R. But this small displacement, by modifying the whole crystal lattice, also changes the refractive index via the photoelastic effect and, through this, the resonance wavelength of the WGM. The total $g_{om}$ can be split into two independent contributions depending on each of these two mechanisms:
\begin{equation}
g_{om}=-\frac{d\omega_0\left(R,\varepsilon\right)}{dx}=-\underbrace{\frac{\partial \omega_0}{\partial R}\frac{\mathrm{d} R}{\mathrm{d} x}}_{\textcolor{blue}{geometric \,g_{om}^{geo}}} - \underbrace{\frac{\partial \omega_0}{\partial \varepsilon}\frac{\mathrm{d}\varepsilon}{\mathrm{d} x}}_{\textcolor{blue}{photoelastic\,  g_{om}^{pe}}}
\label{Eqgomseparationtwoterms}
\end{equation}
where $\varepsilon$ is the material's permittivity, which is no longer necessarily isotropic nor homogeneous inside the disk under stress.
The photoelastic contribution $g_{om}^{pe}$  is obviously unique to resonators where light is confined inside matter, like semiconductor disks, silica toroids or spheres and photonic crystal slabs, and would not appear in an empty  Fabry-Perot optomechanical cavity. For this reason, it has been little considered in the early optomechanics literature \cite{bahl2012observation, favero2012optomechanics, rolland2012acousto, chan2012optimized}.

Note that assuming a purely radial mechanical displacement with maximal amplitude exactly at the periphery of the disk, and assuming the separability of the in-plane and out-of-plane components of the electric field,  the geometric optomechanical coupling in a disk resonator of radius R takes the exact simple form $g_{om}^{geo}=\omega_0/R$ \cite{ding2010high}. 

\section{Radiation pressure in an optomechanical disk resonator}
\label{Sectionradiationpressure}

We  first calculate the radiation pressure exerted by photons confined by total internal reflection inside a circular disk resonator in two different ways: 1) through simple analytical energy and momentum conservation arguments; 2) by 3D Finite Element Method (FEM) computations of the Maxwell Stress Tensor (MST). Our analytical approach provides original helpful formula for whispering gallery optomechanics. While both approaches fittingly yield consistent results, each provides specific insights into the radiation pressure mechanism.
\subsection{Analytical approach}
\subsubsection{Energy conservation argument}

The stored electromagnetic energy in the closed resonator is given by:
\begin{equation}
E= N_{ph} \hbar \omega_0
\label{EMenergyEq}
\end{equation}
with N$_{ph}$ the number of stored photons in the resonator and $\omega_0$ the photon's angular frequency. A small mechanical displacement of the disk $\Delta x$ leads to a change in the photon angular frequency $\Delta\omega_0$ and  stored energy $\Delta E=N_{ph} \hbar \Delta \omega_0$.  Therefore the force associated to this work reads:
\begin{equation}
F=  - \frac{\Delta E}{\Delta x}=-N_{ph}\hbar \frac{\Delta \omega_0}{\Delta x}= N_{ph}\hbar g_{om}
\label{Eqgomforce}
\end{equation}
Using Eq. \ref{Eqgomseparationtwoterms}, we split the total force F into two distinct contributions, linked to radiation pressure  $F_{rp}$ and electrostriction $F_{es}$:
\begin{equation}
F_{rp}=N_{ph}\hbar g_{om}^{geo}\quad \mathrm{and} \quad  F_{es}=N_{ph}\hbar g_{om}^{pe}
\end{equation}
\subsubsection{Momentum conservation argument}
\label{sectionmomentumconservationarguments}
Let us consider the radiation pressure exerted on the outer boundary of a disk resonator by a confined photon, through momentum conservation arguments.
In free space, the momentum associated with a photon of wavelength $\lambda_0$ is $\hbar k_0$, with $k_0=2\pi/\lambda_0$  the free space wavenumber.
When this photon impinges on a rigid mirror with orthogonal incidence, and is perfectly reflected, conservation of momentum dictates that the mirror receives $2\hbar k_0$ momentum.  We now wish to describe how much momentum is transferred  to a circular resonator by a photon confined by total internal reflection, as this photon performs a round-trip. 
Using ray optics considerations, a photon confined inside a regular cavity with $n$  sidewalls will strike the sidewalls $n$ times per round-trip at an angle of $\pi/n$, each time transferring radially a momentum $2\hbar k\,\sin\left(\pi/n\right)$ (Fig. \ref{figintro} (c)).
The radial momentum transfer as a photon completes one round trip is the limit:
\begin{equation}
2\hbar k\lim_{n \to \infty} n\, \sin(\pi/n)= 2 \pi \hbar k
\label{Eqmomentumperrt}
\end{equation}
where $k$ is used instead of $k_{0}$ as we now consider the case of a photon confined inside a dielectric medium. Note the difference with the often encountered 4$\hbar k$ expression stemming from the Fabry-Perot case.
The associated radial force per photon is the momentum transfer per round-trip (Eq. \ref{Eqmomentumperrt}) divided by the cavity round-trip time $\tau_{rt}$ and is written for a disk of radius $R$:
\begin{equation}
F=\frac{dP}{dt}=\frac{2 \pi \hbar k}{\underbrace{2 \pi R n_{\mathrm{eff}}/c}_{\textcolor{blue}{\tau_{rt}}}} =\frac{\hbar k c}{n_{\mathrm{eff}} R}
\end{equation}
with $c$ the speed of light in vacuum. Provided we write the photon momentum $\hbar k$ in a material of refractive index $n_{\mathrm{eff}}$ as $\hbar k=\hbar k_0\,n_{\mathrm{eff}}$  (Minkowski formulation for the photon momentum  in a dielectric  \cite{barnett2010resolution}) and use the geometrical expression $g_{om}^{geo}=\omega_0/R$ for a purely radial displacement of the disk, the radial force $F_{rp}$ exerted by N$_{ph}$ photons takes the simple form:
\begin{equation}
F_{rp}=N_{ph} \frac{\hbar\,k_0\,c}{R} = N_{ph}\,\hbar\, g_{om}^{geo}
\label{Eqradforce}
\end{equation}
which is consistent with what was obtained through energy conservation (Eq. \ref{Eqgomforce}). The radiation pressure $P_{rp}$ exerted on the disk resonator's vertical outer boundary of surface $S=2\pi R\, h$ is:
\begin{equation}
P_{rp}=F_{rp}/S=N_{ph}\times\frac{\hbar k_0 c}{2 \pi  R^2 h}=N_{ph} \times\frac{\hbar c}{\lambda_0 \, R^2 \,h \,  }
\label{Eqradpressure}
\end{equation}
\begin{table}
\centering
\begin{tabular}{l c c c }
\hline
\hline
\textbf{Parameter} & \textbf{Name} & \textbf{Unit}  & \textbf{Value}\\
\hline
\hline
Disk weight & P$_{disk}$ & N & 5.2$\cdot 10^{-14}$ \\
\hline
Radiation pressure force per photon & $F_{rp}/N_{ph}$ & N & 1.5$\cdot 10^{-13}$ \\
\hline
Radiation pressure per photon & $P_{rp}/N_{ph}$ & Pa & 7.5$\cdot 10^{-2}$\\
\hline
\end{tabular}
\caption{Radiation pressure values for a 1 $\mu$m radius, 320 nm thick GaAs disk resonator and $\lambda_0$=1.32 $\mu$m wavelength light.}
\label{Tableopticalparams}
\end{table}
The $\frac{1}{R^2 h}$ dependency in Eq. \ref{Eqradpressure} illustrates the benefit of using small-diameter thin disk resonators. Since both force and pressure exerted by the stored photons are independent of the refractive index of the resonator material,  the benefit of  using high refractive index materials appears only through the reduced disk radii feasible before incurring significant bending losses.
Numerical estimates of $F_{rp}$ and $P_{rp}$ for a 1 $\mu$m radius disk are provided in Table \ref{Tableopticalparams}. The remarkable optomechanical properties of these small resonators are highlighted by the fact that the radiation force $F_{rp}$ exerted on the disk's outer boundary by \emph{a single photon}  is  larger than the disk's own weight.
\subsection{Numerical approach}
\label{subsectionmaxwellstresstensor}
In this section we estimate the magnitude of the radial radiation pressure per confined photon by computing the spatially dependent Maxwell stress tensor (MST) \cite{electrodynamics1998jd}. In a dielectric medium of relative permittivity $\varepsilon_{r}\left(r,z\right)$ and permeability $\mu_r$, the $ij$ components of the MST  are given by:
\begin{equation}
T_{ij}=\varepsilon_0\,\varepsilon_{r}(r,z)\left[E_iE_j-\frac{1}{2}\,\delta_{ij}\,|E|^2\right]+ \mu_0\, \mu_{r}\left[H_iH_j-\frac{1}{2}\,\delta_{ij}\,|H|^2\right]
\label{EqMST}
\end{equation}
Here $\varepsilon_0=8.85\cdot 10^{-12}$ F$\cdot$m$^{-1}$ and $\mu_0=4\pi\cdot 10^{-7}$ H$\cdot$m$^{-1}$ are the vacuum permittivity and permeability, $\delta_{ij}$ is Kronecker's delta and $E_i \, (H_i)$ is the $i$th electric (magnetic) field component. In the following we will take $\varepsilon_{r}\left(r,z\right)=n^2 \in  \mathbb{R} $ and $\mu_r$=1 inside the GaAs, and $\varepsilon_r$=$\mu_r$=1 in the surrounding air. With the choice of notations of Eq. \ref{EqMST}, the radiation pressure induced stress $\sigma_{ij}^{rp}$ (applied on the face normal to the  $i$ direction along the $j$ direction) is expressed as a function of the MST element T$_{ij}$ as $\sigma_{ij}^{rp}=-T_{ij}$. 
While this approach allows for computing both normal ($\sigma_{ii}$) and shear ($\sigma_{ij}$ with $i\neq j$) stresses, in the following we focus only on normal stresses, as these are the ones producing work when coupled to the radial displacement of a  mechanical Radial Breathing Mode (RBM). 
Since the disk cannot respond mechanically to rapidly varying forces at optical frequencies (10$^{14}$ Hz range), we compute the time averaged value of the radial stress over an optical cycle.

\begin{figure}
\centering
\includegraphics[width=\textwidth]{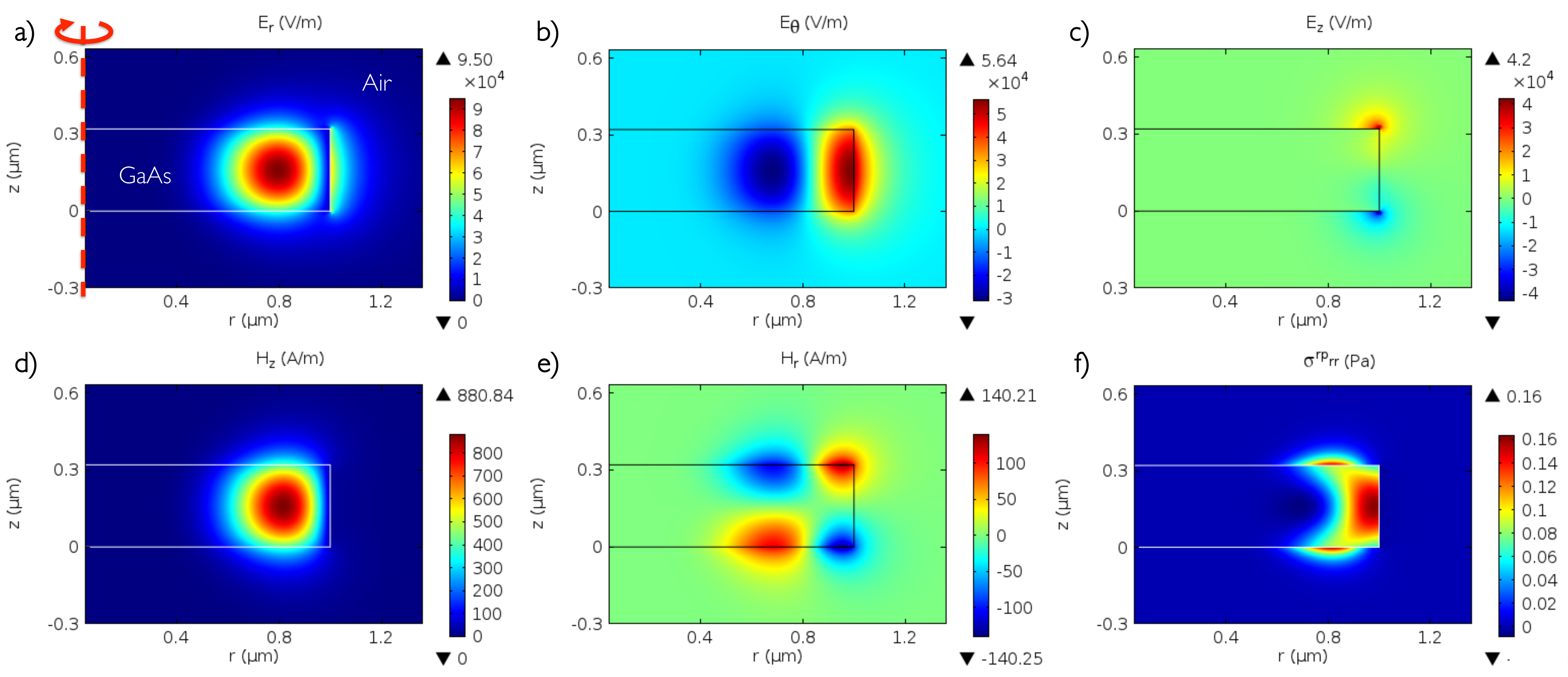}
\caption{2D axi-symmetric FEM modeling of the normal radial `radiation pressure' stress $\sigma_{rr}^{rp}$ in a 320 nm thick and 1 $\mu$m radius GaAs WGM disk resonator. The considered WGM is a (p=1, m=10). The solid lines show the boundary of the two computational domains: the GaAs disk and the surrounding air. 2D axi-symmetric cross sections are shown here, the whole disk is obtained by revolving around the \textit{z}  axial symmetry axis (dashed red line). The AlGaAs pedestal, being sufficiently remote from the optical field, is not included in the simulation. Images (a) through (e) show the computed electric and magnetic field cross-sections, normalized such that the total electromagnetic energy in the resonator is equal to the energy of one photon. (a) $E_r$ (b) $E_\theta$ (c) $E_z$ (d) $H_z$ (e) $H_r$. Since the simulated WGM is TE, the in-plane electric field  and out-of-plane magnetic field components $E_r$, $E_\theta$ and $H_z$ are dominant. (f) Normal radial stress exerted by a confined photon $\sigma_{rr}^{rp}$.  The optically induced stress is largest near the outer boundary of the disk resonator, where most of the electromagnetic energy is located.}
\label{maxwellstresstensorcomsolfig}
\end{figure}

To calculate the radial radiation pressure due to a photon confined in the resonator in a specific WGM, we first perform a FEM simulation of the desired WGM. 
(Throughout this paper -unless mentioned otherwise- we will be considering Transverse Electric (TE) WGMs, with radial order p=1 and a resonance wavelength $\lambda_0\simeq 1.3 \mu$m).
This simulation provides the electric and magnetic field components needed to compute Eq. \ref{EqMST}.  The main field components are plotted in Figure \ref{maxwellstresstensorcomsolfig} a, b, c, d, and e. Next, the value of the time-averaged normal radial stress  $\sigma_{rr}^{rp}=-T_{rr}$ is calculated at every point in space along the $rz$ cross-section (see Fig. \ref{maxwellstresstensorcomsolfig}, f). The normal radial stress is largest near the outer edge of the disk, where the light is confined. From the local stress we can infer a local volume force (force per unit volume) $\mathcal{F}$ via the relation:
\begin{equation}
\mathcal{F}^{rp}_j=-\partial_i \sigma_{ij}^{rp}=\partial_i T_{ij}
\label{Eqvolumeforce}
\end{equation}
The spatial distribution of the radiation volume force  $\mathcal{F}^{rp}_r$ is maximal right at the discontinuous dielectric interface (at r=1 $\mu$m), lending some degree of support to the previously used image of the photon as a particle exerting a force as it bounces off the resonator sidewalls. In this image, the photon is ``pushing on the boundary''. In order to quantitatively compare the results of the analytical approach, which considers a radial force applied to the disk boundary, with the MST approach, which provides radial, azimuthal and axial stresses distributed throughout the disk resonator, the associated g$_{om}^{geo}$ must be computed. This will be done in section \ref{optomechanicalcouplingchapter4section}.

\section{Electrostriction in an optomechanical disk resonator}
\label{Sectionelectrostrictive}
Electrostriction is a mechanism whereby electric fields induce strain within a material. It differs from piezoelectricity in that the induced strain is proportional to the square of the electric field, and not to the electric field. Since the electric fields we consider are rapidly oscillating at optical frequencies, the time averaged piezoelectric strain shall be zero, while the time averaged electrostrictive strain contribution remains. Electrostrictive stresses scale with the fourth power of the dielectric refractive index, making them of significant importance for high refractive index materials such as silicon and GaAs (for which n $\geq$ 3.3 at telecom wavelengths) \cite{rakich2010tailoring}.
The electrostrictively induced stress can be expressed in terms of the material photoelastic tensor $p_{ijkl}$ \cite{feldman1975relations}, which links a material strain $S_{ij}$ to a change in the material's inverse dielectric tensor $\varepsilon_{ij}^{-1}$:
\begin{equation}
\varepsilon_{ij}^{-1}\left(S_{kl}\right)=\varepsilon_{ij}^{-1}+\Delta\left(\varepsilon_{ij}^{-1}\right)=\varepsilon_{ij}^{-1}+p_{ijkl}S_{kl}
\label{Eqphotoelasticity}
\end{equation}
The photoelastic tensor has $3^4=81$ elements, that reduce to only 3 independent coefficients for cubic crystals such as GaAs \cite{newnham2004properties}. These three parameters  are p$_{11}$, p$_{12}$ and p$_{44}$, written here in contracted notation, where 11$\rightarrow$ 1; 22 $\rightarrow$ 2; 33 $\rightarrow$ 3; 23, 32 $\rightarrow$ 4; 31, 13 $\rightarrow$ 5; 12, 21 $\rightarrow$ 6.
Using this definition, Eq. \ref{Eqelectrostrictionphotoelasticity} links the electrostrictively induced stresses $\sigma^{es}$ to the electric field components in the following way \cite{feldman1975relations}:
\begin{equation}
\begin{pmatrix}
\sigma_{rr}^{es}\\
\sigma_{\theta\theta}^{es} \\
\sigma_{zz}^{es} \\
\sigma_{\theta z}^{es}=\sigma_{z\theta}^{es}\\
\sigma_{rz}^{es}=\sigma_{zr}^{es} \\
\sigma_{r\theta}^{es}=\sigma_{\theta r}^{es}\\
\end{pmatrix}
= -\frac{1}{2}\varepsilon_0\,n^4
\underbrace{\begin{pmatrix}
p_{11} & p_{12} & p_{12} & 0 & 0 & 0 \\
p_{12} & p_{11} & p_{12} & 0 & 0 & 0 \\
p_{12} & p_{12} & p_{11} & 0 & 0 & 0 \\
0 & 0 & 0 & p_{44} & 0 & 0 \\
0 & 0 & 0 & 0 & p_{44} & 0 \\
0 & 0 & 0  & 0 & 0 & p_{44}\\
\end{pmatrix}}_{\textcolor{blue}{\mathrm{photoelastic\, tensor}}}
\begin{pmatrix}
E_r^2 \\
E_\theta^2 \\
E_z^2 \\
E_\theta\,E_z \\
E_r\,E_z \\
E_r\,E_\theta\\
\end{pmatrix}
\label{Eqelectrostrictionphotoelasticity}
\end{equation}

\begin{figure}
\centering
\includegraphics[width=.7\textwidth]{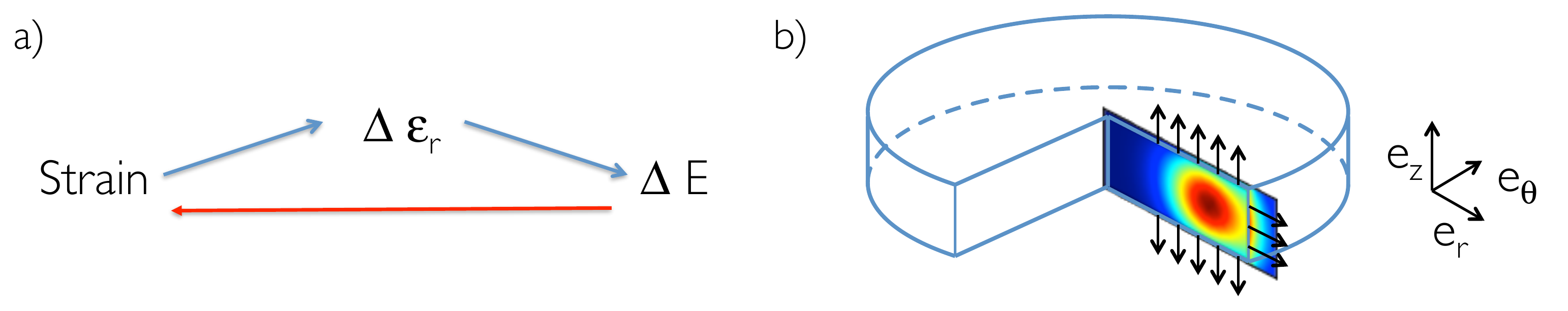}
\caption{(a) Illustration of the link between photoelasticity and electrostriction. A strain leads to change in refractive index (photoelasticity) which itself leads to a change in the stored electric energy. Electrostriction is the converse mechanism (red arrow), whereby electric fields (stored energy) induce strain in the material. (b) Schematic illustrating the direction of electrostrictive and radiation pressure forces acting on a GaAs disk resonator due to photons confined in a WGM (black arrows), and represented in the cross section plane over which the stress and volume force are plotted in Figs. \ref{maxwellstresstensorcomsolfig} and \ref{electrostricitvestressfig}.  }
\label{electrostrictionexplanationfig}
\end{figure}

The value of the three photoelastic coefficients for GaAs are provided in Table \ref{Tablephotoelasticparams}. The relation between electrostriction and photoelasticity is seen by considering a disk resonator suddenly subject to strain. The strain leads to a change in the material's permittivity $\Delta\varepsilon$, via the photoelastic properties. Provided some electric energy was stored in the disk at the time, this stored energy (proportional to $\varepsilon$ E$^2$) changes due to the change in permittivity $\Delta\varepsilon$.  This change in energy can be seen as the work of the electrostrictive force during the displacement. (A more complete version of this argument is developed in \cite{feldman1975relations}, see Fig. \ref{electrostrictionexplanationfig} a). 

\begin{figure}
\centering
\includegraphics[width=.75\textwidth]{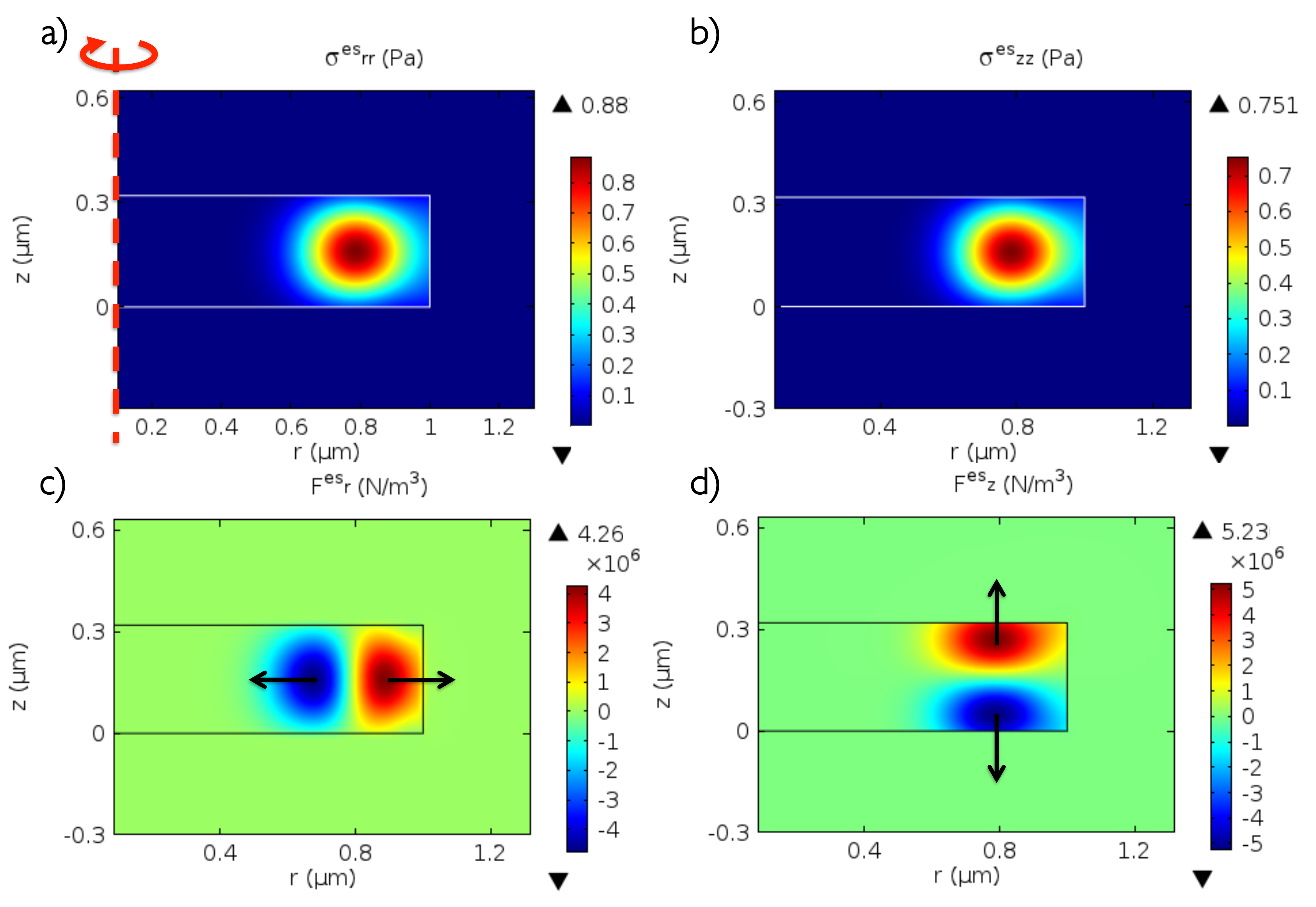}
\caption{2D axi-symmetric FEM modeling of the electrostrictive stress and volume force. 
(a) and (b) show respectively the $rz$ cross-section of the radial $\sigma_{rr}^{es}$ and axial $\sigma_{zz}^{es}$ electrostrictive stress distributions. The azimuthal normal stress $\sigma_{\theta\theta}^{es}$ (not shown here) is of comparable magnitude. (c) and (d) plot the associated radial and axial volume force distributions. Black arrows indicate the overall direction these forces point in. }
\label{electrostricitvestressfig}
\end{figure}

\begin{table}
\centering
\begin{tabular}{l c c c c c}
\hline
\hline
Material & Wavelength ($\mu$m) & $p_{11}$ & $p_{12}$ & $p_{44}$ & Reference\\
\hline
\hline
GaAs & 1.15 & -0.165 & -0.140 & -0.072 & \cite{lide2012crc}  \\
\hline
Si & 3.39 & -0.09 & +0.017 & -0.051 &  \cite{biegelsen1974photoelastic} \\
\hline
\end{tabular}
\caption{Photoelastic material parameters for GaAs,  and silicon (Si) for comparison. The photoelastic coefficients vary little for wavelengths with energies well below the material bandgap \cite{raynolds1995strain}.}
\label{Tablephotoelasticparams}
\end{table}
Looking at Eq. \ref{Eqelectrostrictionphotoelasticity} and the values of the photoelastic coefficients in Table \ref{Tablephotoelasticparams}, it appears in the case of a WGM that the electrostrictively induced normal stresses are significantly larger than the shear stresses. We will focus here for brevity  just on the radial $\sigma^{es}_{rr}$ and axial $\sigma^{es}_{zz}$ normal stress components:

\begin{equation}
\begin{aligned}
\sigma_{rr}^{es}&=-\frac{1}{2}\varepsilon_0\,n^4\left[p_{11}|E_r|^2+p_{12}\left(|E_\theta|^2+|E_z|^2\right)\right] \\
\sigma_{zz}^{es}&=-\frac{1}{2}\varepsilon_0\,n^4\left[p_{11}|E_z|^2+p_{12}\left(|E_r|^2+|E_\theta|^2\right)\right]
\end{aligned}
\label{Eqnormalaxialstresselectrostrictive}
\end{equation}
Figure \ref{electrostricitvestressfig} (a) and (b) show the value of $\sigma^{es}_{rr}$ and $\sigma^{es}_{zz}$ due to a single photon confined in the p=1, m=10 WGM of a 1 $\mu$m radius GaAs disk resonator of thickness 320 nm, already considered in section 2. Figure \ref{electrostricitvestressfig} (c) and (d) represent the associated volume force for both these stresses, where $ \mathcal{F}^{es}_r=-\partial_r \sigma_{rr}^{es}$ and $ \mathcal{F}^{es}_z=-\partial_z \sigma_{zz}^{es}$. The black arrows show the net direction these forces are pointing in. We see here that the electrostrictive force pushes outwards in both the radial and vertical $z$ directions, adding constructively to the radiation pressure force.
The fact that electrostriction and radiation pressure add up constructively as they do here is not true for all materials and geometries. As we can see from Eq. \ref{Eqnormalaxialstresselectrostrictive}, 
since the p$_{ij}$ are negative for GaAs, all electrostrictive stresses are positive,  and confined photons tend to expand the material in all directions. However this would not be the case for silicon disk resonators or waveguides, as the coefficients $p_{11}$ and $p_{12}$ are of different sign and significantly different magnitude.
\section{Optomechanical coupling in GaAs disks}
\label{optomechanicalcouplingchapter4section}
\subsection{Geometric contribution g$_{om}^{geo}$}
\label{Subsectiongeometricgom}
\begin{figure}
\centering
\includegraphics[width=\textwidth]{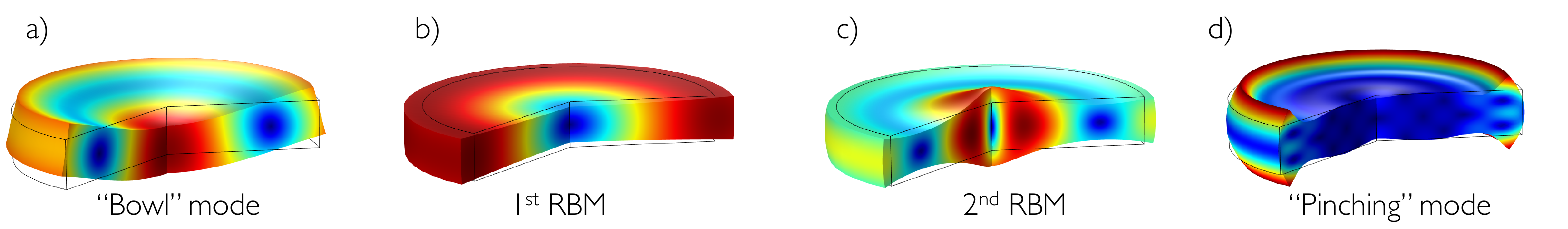}
\caption{(a) through (d): Displacement profile for the four mechanical eigenmodes listed in Table \ref{Tablegom}, with exaggerated deformation. The surface color code illustrates the total displacement, with red as maximum and blue as minimum. }
\label{gomcalculationfig}
\end{figure}

Reference \cite{johnson2002perturbation} provides a perturbation theory for Maxwell's equations in the case of shifting material boundaries. This theory can be applied to determine the frequency shift of an optical WGM to an arbitrary mechanical deformation of the confining dielectric disk. Following this method, the term $g_{om}^{geo}$ is calculated as a surface integral of the unperturbed optical fields over the perturbed dielectric interface:
\begin{equation}
g_{om}^{geo}=\frac{\omega_0}{4}\iint_{\mathrm{disk}} \left(\vec{q}\cdot\vec{n}\right)\left[\Delta\varepsilon_{12} |\vec{e_{\parallel}}|^2-\Delta\left(\varepsilon_{12}^{-1}\right)|\vec{d_{\perp}}|^2\right]\mathrm{d}A
\label{Eqgomperturbation}
\end{equation}
Here $\vec{q}$ and $\vec{n}$ are respectively the normalized mechanical displacement vector and surface normal vector. $\vec{e_{\parallel}}$  (resp. $\vec{d_{\perp}}$) is the parallel (orthogonal) component to the surface of the electric field (electric displacement field). $\vec{q}$ and $\vec{e}$ are normalized such that max$|\vec{q}|$=1 and $\frac{1}{2}\int\varepsilon|e|^2 \mathrm{d}V=1$.
$\Delta \varepsilon_{12}=\varepsilon_1-\varepsilon_2$ is the difference in permittivity between the materials on either side of the boundary and   $\Delta\left(\varepsilon_{12}^{-1}\right)$=$\varepsilon_1^{-1}-\varepsilon_2^{-1}$.
Here we are only interested in the geometric contribution to the g$_{om}$, so $\varepsilon_1$ is simply n$^2$ over the entire disk, while $\varepsilon_2$=1.  $g_{om}^{geo}$ is computed from Eq. \ref{Eqgomperturbation} using a FEM simulation software (COMSOL Multiphysics). The results for the four mechanical modes shown in Fig. \ref{gomcalculationfig} are summarized in Table \ref{Tablegom}.
For a given mechanical mode, the displacement of every point of the disk is spatially non uniform and $g^{geo}_{om}=-\frac{\mathrm{d}\omega_0^{geo}}{\mathrm{d}x}$ is therefore dependent on the somewhat arbitrary choice of the reduction point which experiences the displacement $dx$. The normalization choice max$|\vec{q}|$=1 in Eq. \ref{Eqgomperturbation} means that the point of maximal displacement is used as reduction point. As evidenced in Table \ref{Tablegom}, different mechanical modes have vastly different values of $g_{om}^{geo}$. Note for instance how the $g_{om}^{geo}$ value for the 1$^{\mathrm{st}}$ RBM is roughly 10 000 times larger than for the out of plane `bowl' mode. This difference illustrates how efficiently each mechanical mode modulates the total cavity length (the $\mathrm{d} R/\mathrm{d} x$ term in Eq. \ref{Eqgomseparationtwoterms})
and confirms that the first RBM is the mechanical mode with the highest $g_{om}^{geo}$.

\begin{table}
\centering
\begin{tabular}{l c c c c}
\hline \hline
Mechanical mode & `bowl' & 1$^{st}$ RBM & 2$^{nd}$ RBM & `pinching'\\
\hline \hline
Frequency & 494 MHz & 1.375 GHz & 3.5 GHz & 5.72 GHz \\
\hline
g$_{om}^{geo}$ (GHz/nm) & 0.11 & 1080 & 412 & 82\\
\hline
g$_{om}^{pe}$ (GHz/nm) & 0 & 984 & 1720 & 231\\
\hline
g$_{om}^{total}$ (GHz/nm) & 0.11 & 2064 & 2132 & 313\\
\hline
x$_{\mathrm{ZPF}}$ (m) & 2.95$\cdot$10$^{-15}$ & 1.23$\cdot$10$^{-15}$ & 1.15$\cdot$10$^{-15}$ & 2.17$\cdot$10$^{-15}$
\\
\hline
g$_{0}^{geo}$ (MHz) & 3.2$\cdot$10$^{-4}$ & 1.33 & 0.474 & 0.18\\
\hline
g$_{0}^{pe}$ (MHz) & 0 & 1.21 & 1.98 & 0.50\\
\hline
g$_{0}^{total}$ (MHz) &  3.2$\cdot$10$^{-4}$ & \textcolor{red}{2.54} & \textcolor{red}{2.45} & 0.68\\
\hline
\end{tabular}
\caption{Comparison between the \emph{geometric} and \emph{photoelastic} optomechanical coupling strengths $g_{om}^{geo}$ and $g_{om}^{pe}$, for four mechanical modes of a 320 nm thick, 1 $\mu$m radius GaAs disk, and a p=1 m=10, $\lambda_0 \simeq 1.3$ $\mu$m WGM, obtained through FEM simulations. The mechanical deformation profiles are shown in Fig \ref{gomcalculationfig}.}
\label{Tablegom}
\end{table}
Due to their extremely miniaturized dimensions,   1 $\mu$m GaAs disks exhibit remarkably large optomechanical coupling, with $g_{om}^{geo}$  reaching over 1 THz/nm in the case of the first RBM. (Here we see that the $\sim$1.1 THz/nm numerically computed value is roughly 20\% below the value provided by the simplified expression $g_{om}=\omega_0/R\simeq 1.4$ THz/nm).
The zero point fluctuations x$_{\mathrm{ZPF}}$ are obtained by equaling the mechanical energy in the resonator to $\hbar \Omega_M /2$, yielding x$_{ZPF}$=1.2$\cdot$10$^{-15}$ m for the first RBM, using the same reduction point as above. The single photon optomechanical coupling strength for this mechanical mode is $g_0$=$g_{om} x_{\mathrm{ZPF}}\simeq 1.3$ MHz. Note that these calculations are carried out without any AlGaAs pedestal under the disk, and are therefore only valid for small pedestal radii ($\geq$ 90 \% undercut). For larger radii the stated mechanical frequencies and g$_{om}^{geo}$ may differ significantly. 
\subsection{Photoelastic contribution g$_{om}^{pe}$}
\label{Subsectionphotoelasticgom}
\begin{figure}
\centering
\includegraphics[width=.9\textwidth]{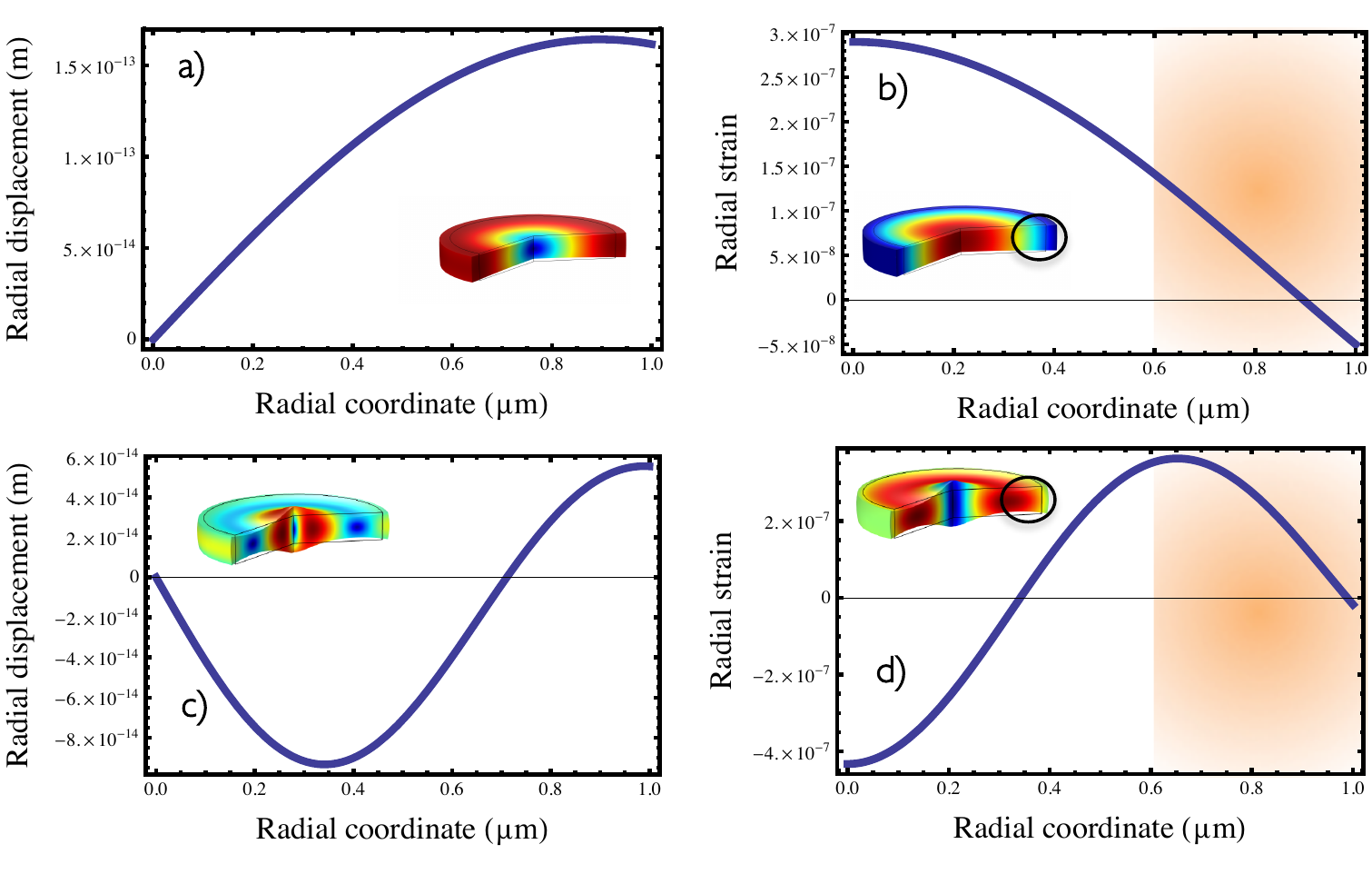}
\caption{Radial displacement and normal radial strain S$_1$=S$_{rr}$ as a function of the radial coordinate, for the first RBM ((a) and (b)) and for the second RBM ((c) and (d)) of a 1$\mu$m radius and 320 nm thick GaAs disk resonator. The values are obtained through FEM modeling, using the approximation of an isotropic Young's modulus for GaAs. The orange highlighted zone between r=0.6 $\mu$m and r=1$\mu$m and the black ring in the inset pictures mark the region of highest electromagnetic energy density for a p=1 WGM. (The displacements are normalized such that for both modes the mechanical energy is equal to $k_B$T at 300 K).}
\label{dispstrain3dfig}
\end{figure}

To compute the photoelastic coupling contribution, first the unperturbed resonance frequency of the desired WGM is obtained through a FEM simulation with uniform and isotropic $\varepsilon$. Second, the desired mechanical eigenmode is solved for in another FEM simulation, which provides the complete deformation profile and strain distributions inside the resonator (we will focus in the following discussion on the first RBM). 
While the radial displacement is zero at the center and maximum near the periphery, the behavior for the normal radial strain is reversed. The normal radial strain S$_{rr}$=S$_1$ (in contracted notation)  is maximal at the center of the disk and changes sign right by the edge of the disk  (this is a normal consequence of the circular geometry), see Fig. \ref{dispstrain3dfig} (a) and (b). The behavior is similar for the normal azimuthal and axial strains S$_2$ and S$_3$, which are of similar magnitude and largest near the center of the disk. The S$_4$ and S$_6$ strain components are zero over the whole disk, while the S$_5$ strain component is roughly three orders of magnitude smaller than S$_{1,2,3}$.
We now use Eq. \ref{Eqphotoelasticity} to relate the strain distribution inside the disk to changes in the dielectric tensor.
Since S$_4$, S$_5$ and S$_6$ are negligible, the off-diagonal terms in the dielectric tensor can be neglected. The dielectric tensor modified by the RBM displacement therefore takes the form:
\begin{equation}
\begin{pmatrix}
\varepsilon_1 & &  \\
& \varepsilon_2 &\\
& & \varepsilon_3 \\
\end{pmatrix}
\;\;\mathrm{with}\;
\begin{cases}
\varepsilon_1=\left(1/n^2+p_{11} S_1+p_{12} S_2 + p_{12} S_3\right)^{-1}\\
\varepsilon_2=\left(1/n^2+p_{12} S_1+p_{11} S_2 + p_{12} S_3\right)^{-1}\\
\varepsilon_3=\left(1/n^2+p_{12} S_1+p_{12} S_2 + p_{11} S_3\right)^{-1}
\end{cases}
\label{Eqmodifieddielectrictensor}
\end{equation}
Note that it is now both anisotropic and dependent upon the position inside the disk resonator.
The problem of finding the new WGM resonance frequency under these conditions is solved through another FEM simulation (with unperturbed geometric boundaries). This provides the photoelastic frequency shift due to the mechanical displacement $dx$. In the linear limit of small $dx$, the procedure leads to the photoelastic optomechanical coupling $g_{om}^{pe}$, which is found to amount to 0.98 THz/nm for the first RBM of the above considered disk and WGM.
This value is remarkably high, considering how inefficient the refractive index modulation is through the first RBM. Indeed in order to maximize the photoelastic  frequency shift \emph{the optical mode should be localized in the region of highest strain}. In the case of the first RBM the radial strain is not only weak but also changes sign right around the area of highest optical energy density (see the highlighted area of Fig. \ref{dispstrain3dfig} b). In contrast, this condition is much better fulfilled for the second order RBM (Fig. \ref{dispstrain3dfig} (c) and (d)). Accordingly, this translates into a remarkably high $g_{om}^{pe}$ of nearly 2 THz/nm for this mechanical mode, see Table \ref{Tablegom}. Because of the reduced geometric coupling for the second RBM, its total optomechanical coupling g$_0^{total}$ is comparable to that of the first RBM, around 2.5 MHz, albeit at a much higher mechanical frequency of 3.5 GHz. Table \ref{Tablegom} summarizes the g$_0$ values for the four considered mechanical modes of Fig. \ref{gomcalculationfig}.

\begin{figure}
\centering
\includegraphics[width=\textwidth]{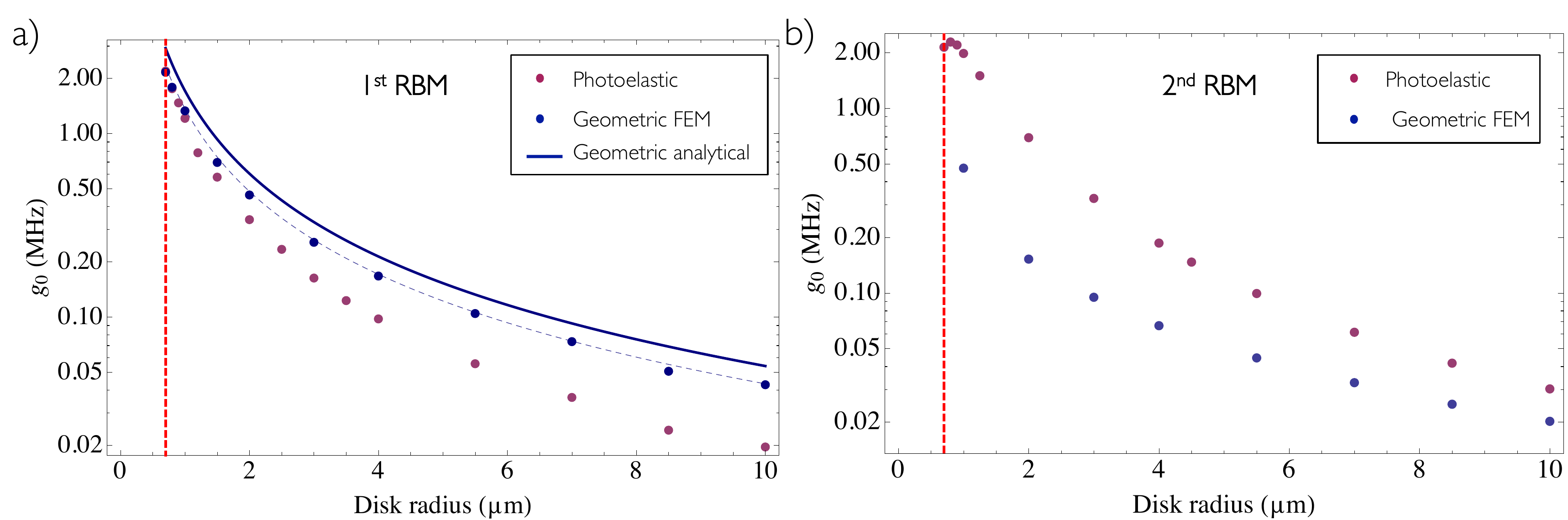}
\caption{Comparison between the geometric and photoelastic optomechanical coupling strength g$_0$ for the TE (p=1) WGM of a GaAs disk of thickness 320 nm, as a function of radius, for the first RBM (a) and second RBM (b), in log scale. The vertical dashed red line represents the radius r=0.7$\mu$m, at which bending losses become limiting at the considered WGM wavelength of 1.3 $\mu$m. For the first RBM, the combined optomechanical coupling g$_{0}^{pe}$+ g$_{0}^{geo}$ reaches 4 MHz for the smallest disks. (The dashed blue line represents the values given by Eq. \ref{Eqg0geo} reduced by 20\%). For the second RBM, photoelasticity is the dominant optomechanical coupling mechanism, with g$_{0}^{pe}$ reaching 2 MHz for R=1 $\mu$m.}
\label{figureg0}
\end{figure}

Figure \ref{figureg0} plots the dependency of g$_0^{pe}$ and g$_0^{geo}$ with disk radius for the first and second RBM.
Each purple (blue) dot corresponds to a distinct FEM simulation of  g$_0^{pe}$ (g$_0^{geo}$), while the solid blue line for g$_0^{geo}$ corresponds to the value given by the following analytical formula:
\begin{equation}
g_0^{geo}=g_{om}^{geo} x_{\mathrm{ZPF}}\simeq  \frac{\omega_0}{R} \sqrt{\frac{\hbar}{2\,m_{\mathrm{eff}}\,\Omega_{M}^P}} \quad \mathrm{with} \quad \Omega_{M}^P=\dfrac{\lambda_P}{R}\,\sqrt{\dfrac{E}{\rho\left(1-\nu^2\right)}}
\label{Eqg0geo}
\end{equation}
Here m$_{\mathrm{eff}}$, $\lambda_P$, $E$, $\rho$ and $\nu$ are respectively the effective mass of the mode, calculated with a reduction point sitting on the disk boundary, a frequency parameter, the Young Modulus, density and Poisson ratio of GaAs, the values of which are provided in Table \ref{Tablelambdap}. $\Omega_{M}^P$ is the mechanical frequency of the RBM of order P  \cite{love2013treatise, onoe1956contour}. We have obtained Eq. \ref{Eqg0geo} through the analytical treatment of a free elastic circular plate. For the first RBM,  Eq. \ref{Eqg0geo} accurately reproduces the trend provided by FEM simulations, but overestimates the coupling by 20\%  because it neglects the out-of-plane component of the mechanical motion. For the second RBM the overestimation is more pronounced, reflecting a larger out-of-plane component of the mechanical mode.\\
Since the effective mass scales with $R^2$, g$_0^{geo}$ scales as $\left(\frac{1}{R}\right)^{3/2}$. Interestingly, g$_0^{pe}$ rises faster than g$_0^{geo}$ with decreasing disk radius. For instance for the 1st RBM (Fig \ref{figureg0} a), g$_0^{pe}$ goes from being two times smaller than g$_0^{geo}$ for disks of radius R=10$\mu$m, before reaching comparable values for 1$\mu$m radius disks.  
For the 2nd RBM (Fig \ref{figureg0} b), the photoelastic coupling is always the dominant coupling mechanism. Note that the maximal photoelastic coupling is reached for $R\simeq 1 \mu$m. Further reducing the disk dimensions reduces the coupling as the optical mode is no longer well localized on the region of highest strain.
 
\begin{table}
\centering
\begin{tabular}{c c c  c c c c c c}
\hline 
\hline
\multicolumn{3}{c}{\textbf{Frequency parameter $\lambda_P$}}& \multicolumn{3}{c}{\textbf{Effective mass ratio}} & \multicolumn{3}{c}{\textbf{GaAs material parameters}}      \\
\hline
\hline
$\lambda_1$ &  $\lambda_2$ & $\lambda_3$   & $m_{\mathrm{eff} 1}/m$ &  $m_{\mathrm{eff} 2}/m$ & $m_{\mathrm{eff} 3}/m$& $\rho$ [kg$\cdot$m$^{-3}$]&$E$ [GPa]& $\nu$\\
\hline
2.055 & 5.391 & 8.573   & 0.786 & 0.969 & 0.988&5317  & 85.9 & 0.31\\
\hline
\end{tabular}
\caption{First three values of the frequency parameter $\lambda_P$ and effective mass ratios for GaAs disk RBMs, and GaAs material parameters used in the calculations. The effective mass ratio is defined as the effective mass associated to  a reduction point on the disk boundary $m_{\mathrm{eff}}$ divided by the disk mass $m$.}
\label{Tablelambdap}
\end{table}

\subsection{Energy considerations}

The link between radiation pressure and boundary deformation and electrostriction and photoelasticity can be understood by looking at the work done by the optical forces during a mechanical displacement.
Incidently, these energy considerations provide an additional way of calculating both the geometric and photoelastic optomechanical couplings. Following Eq. \ref{Eqgomforce}, we can write a generalized expression:
\begin{equation}
g_{om}=\frac{\frac{1}{2}\,\iiint_{\mathrm{disk}}\,\Sigma\sigma_{ij} S_{ij}\;\; \mathrm{d}V}{\Delta x\;\hbar}
\end{equation}
Here the numerator corresponds to the work produced by the optical stress due to \emph{a single} confined photon, during the displacement $\Delta x$, in the case of a linear elastic solid starting at rest \cite{saada2009elasticity}. The S$_{ij}$ are the mechanical strain components resulting from the displacement $\Delta x$, and the $\sigma_{ij}$ are the radiation pressure or electrostrictive stress components described respectively in Eqs. \ref{EqMST} and \ref{Eqelectrostrictionphotoelasticity}.
This formulation and the method discussed in \ref{Subsectionphotoelasticgom} yield values in very good agreement, within less than 1 \% difference.
Note that both for radiation pressure and electrostriction, in the case of the 1st RBM at least, a large part of the work is done by the optically induced azimuthal stress $\sigma_{\theta\theta}$. Furthermore, for the same mechanical mode, the larger axial stress $\sigma_{zz}$ in the case of electrostriction produces negative work as the disk expands in the radial direction but contracts in the axial direction. These considerations shed light on two seemingly contradictory observations. On one hand the photoelastic coupling $g_{0}^{pe}$ is slightly smaller than the geometric coupling $g_{0}^{geo}$ for the first RBM, on the other hand the radial stress per photon is several times larger for electrostriction than radiation pressure. 
As a consequence, even though the movement  of the 1st RBM is predominantly radial,  the full picture of optomechanical coupling  can not be obtained looking solely at the forces exerted in the radial direction.
\subsection{Discussion}
We show that the second order RBM is an interesting mechanical mode thanks to its large total optomechanical coupling and high mechanical frequency. While this type of mode tendentially has a lower mechanical Q due to larger mechanical coupling to the pedestal, its anchoring losses could be overcome with a carefully engineered pedestal geometry \cite{nguyen2013ultrahigh}.

We verify that both the geometric and photoelastic coupling magnitudes are comparable when considering a transverse magnetic (TM) WGM instead of a TE WGM, with values varying by less than 20 \%. We focused here on p=1 WGM, as these are  the modes with the highest radiative optical Qs \cite{ding2010high}. When considering different WGMs, the same rule of thumb remains: in order to maximize the photoelastic  optomechanical coupling, the regions of high electromagnetic energy should be co-localized with regions of high mechanical  strain.

For comparison the photoelastic optomechanical coupling has been computed on Si disks of identical dimensions using the photoelastic parameters of Table \ref{Tablephotoelasticparams}. The obtained g$_0^{pe}$ for the 1st RBM is roughly three times lower than for GaAs, notably because of the reduced photoelastic coefficients of Si, but should nevertheless not be neglected.

Finally recent work investigating the optomechanical coupling in distributed Bragg reflector GaAs/AlAs vertical cavities \cite{fainstein2013strong} shows these geometries are also well suited to take advantage of the photoelastic coupling mechanism, thanks to an efficient overlap between the optical field and strain maxima resulting in values of g$_{om}^{pe}$ reaching several THz/nm. 

\section{Conclusion}
%

We investigated the magnitude of the optical forces due to confined photons in GaAs semiconductor optomechanical disk resonators, successively addressing the case of radiation pressure and electrostriction. We showed these forces add up constructively in the case of GaAs disks. Next, we provided a comparison between the photoelastic and geometric optomechanical coupling for various modes of a GaAs disk, and the scaling of these couplings with disk radius. An interpretation of this coupling in terms of the work done by the optical forces during a mechanical displacement is proposed and numerically verified, leading to an additional estimation of g$_0$. Photoelasticity provides an efficient tool when designing structures for optomechanical applications. The large photoelastic coupling in GaAs underscores the strength of this material for optomechanical applications, in complement with other coupling mechanisms proposed in GaAs membranes and cantilevers \cite{usami2012optical, okamoto2011vibration}.

\section*{Acknowledgements}
This work is supported by the French ANR through the NOMADE project and by the ERC through the GANOMS project. The authors would like to thank Bernard Jusserand for insightful discussions on photoelasticity.
\end{document}